\documentstyle[prb,twocolumn,aps]{revtex}

\input{epsf}

\begin{document}

\tolerance 10000

\twocolumn[\hsize\textwidth\columnwidth\hsize\csname %
@twocolumnfalse\endcsname

\draft

\title{Balanced Branching in Transcription Termination}

\author{K. J. Harrington and R. B. Laughlin}

\address{Department of Physics, Stanford University,
        Stanford, California 94305}

\author{S. Liang}

\address{NASA Ames Research Center, Moffett Field, California 94035}

\date{\today}
\maketitle
\widetext

\begin{abstract}
\begin{center}
 
\parbox{14cm}{The theory of stochastic transcription termination based on
free-energy competition \cite{standard} requires two or more reaction
rates to be delicately balanced over a wide range of physical conditions.
A large body of work on glasses and large molecules suggests that this
should be impossible in such a large system in the absence of a new
organizing principle of matter.  We review the experimental literature of
termination and find no evidence for such a principle but many troubling
inconsistencies, most notably anomalous memory effects.  These suggest
that termination has a deterministic component and may conceivably be not
stochastic at all.  We find that a key experiment by Wilson and
von Hippel \cite{wilson} allegedly refuting deterministic termination was
an incorrectly analyzed regulatory effect of Mg$^{2+}$ binding.}

\end{center}
\end{abstract}

\pacs{
\hspace{1.9cm}
}
]

\narrowtext

\section{Introduction}

The branching ratio of the termination process in gene transcription is
balanced. In the case most thoroughly studied, $\rho$-independent
termination in procaryotes, conventional gel experiments performed {\it in
vitro} find a fraction $P$ of elongating RNA polymerase reading through
the termination sequence with $| \ln (1/P - 1) | < 4$ essentially always,
even though $P$ is different for different terminators and can be made to
exhibit order-1 changes by perturbing the environment. This effect is
astonishing from the standpoint of microscopic physics because a
stochastic decision to read through or not requires a competition of
transition rates - quantities of inverse time - that must be nearly equal
for the branching to be balanced.  RNA polymerase, however, is more the
size of a glass simulation than a small molecule and thus possesses a
broad spectrum of natural time scales spanning many decades. Without some
physical reason for a particular scale to be preferred, rate competition
ought to have been severely unbalanced, meaning that one event occurs
essentially always and the other never. Balanced branching in termination
has been implicated in transcription regulation in a few cases,
\cite{landick} but its broader significance, especially its robustness, is
still a mystery.

In this paper we examine the experimental facts relevant to the physical
nature of termination with the goal of determining what, if any, principle
selects the time scale for stochastic rate balance.  Our conclusion is
both surprising and unsettling.  We find no evidence for such a principle,
but glaring weaknesses in the case for stochasticity and a large body of
unexplained experimental results pointing to a termination decision that
is partially deterministic.  In light of the inaccessability of systems
this large to ab-initio computation we conclude that transcription
termination is a fundamentally unsolved problem in mesoscopic physics and
an ideal target for the emerging techniques of nanoscience.

\section{Termination Efficiency}

The simplest termination sequences are the $\rho$-independent terminators
of procaryotes, which are capable of causing polymerase to terminate {\it
in vitro} in the absence of the $\rho$ protein factor.  A representative
sampling of these is reproduced in Table I.  This differs from lists that
have appeared in the literature before \cite{brendel,thermes} by having
 been rechecked against the fully-sequenced genome \cite{genome} and
expunged of ``theoretical'' terminators identified only by computer
search.  They conform for the most part to the motif of a palindrome of
typically 10 base pairs followed by a short poly-T stretch, although there
is tremendous variety in the length and composition of the palindrome,
variation in the length of the poly-T stretch, and occasional extension of
the palindrome to include the poly-T stretch.  This enormous variability
contrasts with the simplicity of stop codons, which terminate protein
synthesis by ribosomes and have no other function.

$\rho$-independent terminators are characterized by ``efficiencies'',
i.e., the fraction of assayed transcripts that terminate.  These rarely
take on extreme values close to 1 or 0 when measured {\it in vitro}. In
cases where a measurement {\it in vivo} exists as well the latter is
usually larger \cite{chamberlin} and is occasionally unmeasurably close to
1. Balanced termination efficiency is commonly observed {\it in vivo} as
well, however. Table II shows results from a particularly careful study
\cite{chamberlin} {\it in vitro} in which termination probabilities in
{\it E. coli} for wild-type terminators, mutant terminators, phage
terminators, \cite{phage} and terminators from {\it S. Boydii} were
measured under identical conditions. Despite the great variety of these
\enlargethispage*{1000pt}
sequences the termination efficiency runs only \linebreak\pagebreak

\twocolumn[\hsize\textwidth\columnwidth\hsize\csname %
@twocolumnfalse\endcsname
\begin{table}
\begin{tabular}{|c|c|c|c|c|}
Sequence\cite{brendel,thermes} & Name & Address \cite{genome} & $\pm$
& Reference \\
\hline
{\tt
CGTTAATCCGCAAATAACGT\underline{AAAAACCCGC}TTCG\underline{GCGGGTTTTT}TTATGGGGGGA}
& rpoC t & 4187152 & + & RNA polymerase operon\cite{rpoct} \\
{\tt
CAGTTTCACCTGATTTACGT\underline{AAAAACCCGC}TTCG\underline{GCGGGTTTTT}GCTTTTGGAGG}
& M1-RNA & 3267812 & - & RNA of RNase P \cite{m1rna} \\
{\tt
CGTACCCCAGCCACATTA\underline{AAAAAGCTCGC}TTCG\underline{GCGAGCTTTTT}GCTTTTCTGCG}
& sup & 0695610 & - & supBC tRNA operon \cite{supbc} \\
{\tt
ACACTAATCGAACCCGGCTCAAAG\underline{ACCCGC}TGCG\underline{GCGGGT}TTTTTTGTCTGTAAT}
& & 1260102 & - & Nucleotide synthesis \cite{sprs} \\
{\tt
AGTAATCTGAAGCAACGT\underline{AAAAAAACCCGCC}CC\underline{GGCGGGTTTTTTT}ATACCCGTA}
& L17 & 3437202 & - & Ribosomal RNA operon \cite{l17} \\
{\tt
TCTCGCTTTGATG\underline{TAACAAAAAACCCCGCC}CC\underline{GGCGGGGTTTTTTGTTA}TCTGCT}
& rpm & 3808820 & - & Ribosome rpm operon \cite{rpm} \\
{\tt
GAGTAAGGTTGCCATTT\underline{GCCCTCCGC}TGCG\underline{GCGGGGGGC}TTTTAACCGGGCAGGA}
& t2 & 3306624 & - & Polynucleotide phosphorylase \cite{t2} \\
{\tt
CGATTGCCTTGTGAA\underline{GCCGGAGCGG}GAGA\underline{CTGCTCCGGC}TTTTTAGTATCTATTC}
& deo t & 4619189 & + & deo operon \cite{deot} \\
{\tt
CGTAAAGAAATCAGATACCC\underline{GCCCGC}CTAATGA\underline{GCGGGC}TTTTTTTTGAACAAAA}
& trp a & 1321015 & - & tryptophan synthesis \cite{trpa} \\
{\tt
GCGCAGTTAATCCCACA\underline{GCCGCCAG}TTCCG\underline{CTGGCGGC}ATTTTAACTTTCTTTAA}
& trp t & 1314395 & - & tryptophan synthesis \cite{trpt} \\
{\tt
AAATCAGGCTGAT\underline{GGCTGGTGACT}TTTT\underline{AGTCACCAGCC}TTTTTGCGCTGTAAGG}
& rplL t & 4178530 & + & Ribosomal proteins L7/L12 \cite{rpllt} \\
{\tt
AGGAAACACAG\underline{AAAAAAGCCCGCAC}CTGACA\underline{GTGCGGGCTTTTTT}TTTCGACCAA}
& thr a & 0000263 & + & threonine operon \cite{thra} \\
{\tt
AGCACGCAGTCAAAC\underline{AAAAAACCCGCGC}CATT\underline{GCGCGGGTTTTTT}TATGCCCGAA}
& leu a & 0083564 & - & leucine synthesis \cite{leua}  \\
{\tt
CCCGTTGATCACCCATTCCC\underline{AGCCCCTC}AATC\underline{GAGGGGCT}TTTTTTTGCCCAGGC}
& pyrBI a & 4469985 & - & pyrimidine synthesis \cite{pyrbia} \\
{\tt
ACACGATTCC\underline{AAAACCCCGCCGG}CGCAAA\underline{CCGGGCGGGGTTTT}TCGTTTAAGCAC}
& ilvB a & 3850449 & - & ilvB operon \cite{ilvba} \\
{\tt
GAAACGGAAAACAGCGCCTG\underline{AAAGCCTCC}CAGT\underline{GGAGGCTTT}TTTTGTATGCGCG}
& pheS a & 1797160 & - & Phenylalanyl-tRNA synthetase \cite{phesa} \\
{\tt
CTTAACGAACTAAG\underline{ACCCCCG}C\underline{ACC}GAAA\underline{GGT}C\underline{CGGGGGT}TTTTTTTGACCTTAA}
& ilvGEDA a & 3948053 & + & ilvGEDA operon \cite{ilvgeda} \\
{\tt
CCGCCCCTGCCAGAAATC\underline{ATCCTTA}GCGAAACG\underline{TAAGGAT}TTTTTTTATCTGAAA}
& rrnC t & 3944645 & + & Ribosomal RNA operon \cite{rrnct} \\
{\tt
CATCAA\underline{ATAAAACAAAAGGC}T\underline{CAGTC}GGAA\underline{GACTG}G\underline{GCCTTTTGTTTTAT}CTGTT}
& rrnD t & 3421006 & + & Ribosomal RNA operon \cite{rrndt} \\
{\tt
TCCGCCACTTATTAAGAAG\underline{CCTCGAG}TTAACG\underline{CTCGAGG}TTTTTTTTCGTCTGTA}
& rrnF (G) t & 0228998 & + & Ribosomal RNA operon \cite{rrnft} \\
{\tt
GCATCGCCAATGTAAATC\underline{CGGCCCGCC}TAT\underline{GGCGGGCCG}TTTTGTATGGAAACCA}
& frdB t & 4376529 & - & Fumarate reductase \cite{frdbt} \\
{\tt
TGAATATTTTAGC\underline{CGCCCCAGTCA}GTAA\underline{TGACTGGGGCG}TTTTTTATTGGGCGAA}
& spot42-RNA & 4047542 & + &  spot42 RNA \cite{spot42t} \\
{\tt
ATTCAGTAAGCAGAA\underline{AGTCAAAAGCCTCCG}AC\underline{CGGAGGCTTTTGACT}ATTACTCA}
& tonB t & 1309824 & + & Membrane protein \cite{tonbt} \\
{\tt
AGAAACAGCAAACAATCC\underline{AAAACGCCGC}GTTCA\underline{GCGGCGTTTT}TTCTGCTTTTCT}
& glnS T & 0707159 & + & Glutaminyl-tRNA synthetase \cite{glnst} \\
{\tt
CTGGCATAAGCCAGTTG\underline{AAAGAGGGAG}CTAGT\underline{CTCCCTCTTT}TCGTTTCAACGCC}
& rplT t & 1797371 & - & Ribosome protein L20 \cite{l20} \\
{\tt
GCATCGCCAATGTAAATC\underline{CGGCCCGCC}TAT\underline{GGCGGGCCG}TTTTGTATGGAAACCA}
& ampC a & 4376529 & - & $\beta$-lactamase \cite{ampla} \\
{\tt
TGCGAAGACGAACAAT\underline{AAGGCC}T\underline{CCC}AAATC\underline{GGG}G\underline{GGCCTT}TTTTATTGATAACA}
& phe a & 2735697 & + & Phenylalanine operon \cite{phea} \\
{\tt
ACGCATGAG\underline{AAAGCCCCCGGAAG}ATCAC\underline{CTTCCGGGGGCTTT}TTTATTGCGCGGT}
& hisG a & 2088121 & + & ATP synthesis \cite{hisga} \\
{\tt
CATCAA\underline{ATAAAACGAAAGGC}T\underline{CAGTC}GAAA\underline{GACTG}G\underline{GCCTTTCGTTTTAT}CTGTT}
& rrnB t$_1$ & 4169333 & + & Ribosomal RNA operon \\
{\tt
GGCATCAAATTAAGCAG\underline{AAGGCCATCC}TGAC\underline{GGATGGCCTT}TTTGCGTTTCTACA}
& rrnB t$_2$ & 4169493 & + & Ribosomal RNA operon \\
{\tt
AAT\underline{TAATGTGAG}TTAG\underline{CTCAC}TC\underline{ATTA}GGCACCCCAGGCTTTACACTTTATGCTT}
& lacI tII & 0365588 & - & Lactose synthesis \cite{lacitii} \\
{\tt
CTT\underline{TT}T\underline{GGCGGAGGGCG}TTG\underline{CGCT}T\underline{CTCCGCC}C\underline{AA}CCTATTTTTACGCGGCGGTG}
& uvrD a & 3995538 & + & DNA helicase II \cite{uvrda} \\
\end{tabular}
\caption{$\rho$-independent terminators in {\it E. coli} taken primarily
         from Brendel et al. \cite{brendel} These are
         oriented in the reading direction and are aligned at the
         poly-T stretch.  The palindrome is underlined.  The beginning and
         end of the selected sequences have no absolute meaning but simply
         follow the convention of d'Aubenton et al. \cite{thermes} The
         address identifies the location in the standard {\it E. coli}
         genome \cite{genome} of the left-most nucleotide in the table.}
\end{table}
]

\begin{table}
\begin{tabular}{|c|c|r|}
Sequence & Name & \% T \\
\hline
{\tt 
\underline{GGCTCAGTC}GAAA\underline{GACTGGGCC}TTTCGT\underline{TTT}AAT} 
& rrnB t$_1$ & 84 $\pm$ 1 \\
{\tt
TCAAA\underline{AGCCTCCG}AC\underline{CGGAGGCT}TTTGA\underline{CT}ATTA} 
& tonB t & 19 $\pm$ 1 \\
{\tt
CC\underline{AGCCCGC}CTAATGA\underline{GCGGGCT}TTTTT\underline{TT}GAAC} 
& trp a & 71 $\pm$ 2 \\
{\tt
CC\underline{AGCCCGC}CTAATGA\underline{GCGGGCT}TT\underline{TG}CAAGGTT} 
& trp a 1419 & 2 $\pm$ 1 \\
{\tt
CCA\underline{GCCCGC}CTAATAA\underline{GCGGGC}TTTTTT\underline{TT}GAAC} 
& trp a L126 & 65 $\pm$ 4 \\
{\tt
CCAGC\underline{CCGC}CTAATAA\underline{GCGG}ACTTTTTT\underline{TT}GAAC} 
& trp a L153 & 8 $\pm$ 4 \\
{\tt
CT\underline{GGCTCACC}TTCG\underline{GGTGGGCC}TTTCTG\underline{CG}TTTA}
& T7T$_e$ & 88 $\pm$ 2 \\
{\tt
\underline{GGC}T\underline{CACC}TTCACG\underline{GGTG}A\underline{GCC}TTTCTT\underline{CG}TTCX}
& T3T$_e$ & 14 $\pm$ 2 \\
{\tt
\underline{GGCCTGC}TGGTAATC\underline{GCAGGCC}TTTTTA\underline{TTT}GGG} 
& tR2 & 49 $\pm$ 4 \\
{\tt
AA\underline{ACCACCGTT}GGT\underline{TAGCGGTGG}TTTTTTGT\underline{TTG}} 
& RNA I & 73 $\pm$ 4 \\
\end{tabular}
\caption{Termination efficiencies measured {\it in vitro}.
\cite{chamberlin} The first 3 terminators are native to {\it E.\ coli}.
These are followed by 3 mutants, 3 phage terminators, \cite{phage} and one
from {\it S. Boydii}. Far-right underlined sequences are termination
zones.}
\end{table}

\begin{table}
\begin{tabular}{|r|c|c|}
\multicolumn{1}{|c|}{Sequence} & Name & \% T \\
\hline
{\tt GTTAATAAC\underline{AGGCCTGC}TGGTAATC\underline{GCAGGCCT}TTTTATT}
& tR2 & 40 \\
{\tt GTTAATAAC\underline{AGGGGACG}TGGTAATC\underline{CGTCCCC}TTTTTATT}
& tR2-6 & 56 \\
{\tt TAATAAC\underline{AGGCCTGGC}TGGTAATC\underline{GCCAGGCCT}TTTTATT}
& tR2-11 & 54 \\
{\tt CCGGGTTAATAAC\underline{AGGCCTGC}TTCG\underline{GCAGGCCT}TTTTATT}
& tR2-12 & 69 \\
{\tt CGGGTTATTAACA\underline{GGCCTC}TGGTAATC\underline{GAGGC}TTTTTATT}
& tR2-13 & 11 \\
{\tt ATAACA\underline{GGGGACG}TGGTAATC\underline{GCCAGCAGGCC}TTTTTATT}
& tR2-14 & 20 \\
{\tt GTT\underline{AATAAAAGGCCTGC}TGGTAATC\underline{GCAGGCCTTTTTATT}}
& tR2-16 & 36 \\
{\tt GGTTCTTCTC\underline{GGCCTGC}TGGTAATC\underline{GCAGGCC}TTTTTATT}
& tR2-17 & 67 \\
\end{tabular}
\caption{Termination efficiencies for modified versions of the phage
        $\lambda$ terminator tR2. \cite{hippel}}
\end{table}

\enlargethispage*{1000pt}
\noindent
from 2\% to 88\%. Many other researchers report similar values for
terminators in {\it E.\ coli} and other bacteria, \cite{gross}
including artificially altered terminators.\cite{lynn}\linebreak\pagebreak

\enlargethispage*{1000pt}

\begin{table}
\begin{tabular}{|c|c|c|c|}
Sequence & Name & rpo+ & rpo203 \\
\hline
{\tt GCAACCG\underline{CTGGGG}AATT\underline{CCCCAG}TTTTCA} 
& trpC 301 & 0 & 20 \\
{\tt AACCG\underline{CTGGCCGG}GAT\underline{CGGCCAG}TTTTCA} 
& trpC 302 & 8 & 35 \\
{\tt C\underline{AGCCGCCAG}TTCCG\underline{CTGGCGGCT}TTTAA} 
& trp t & 25 & 45 \\
{\tt ACC\underline{AGCCCGC}CTAATGA\underline{GCGGGCT}TTTGC} 
& trp a 1419 & 3 & 35 \\
{\tt \underline{CAGCCCGC}CTAATGA\underline{GCGGGCTG}TTTTTT} 
& trp a 135 & 65 & 80 \\
\end{tabular}
\caption{Termination efficiences for wild-type {\it E. coli} polymerase
        (rpo+) and mutant polymerase (rpo203). \cite{platt} {\it trp~t} is
	native to the genome.  The rest are either mutants or synthetic.}
\end{table}

\noindent 
The results in Tables III and IV show balanced termination for modified
versions of the phage terminator {\it tR2} \cite{hippel} and for mutant
polymerase.  \cite{platt} This also makes order-1 changes to the
efficiencies themselves.  Similar effects were reported by other
researchers \cite{gross,fisher} with different mutant polymerases. 
Modifications up to 20 base pairs upstream and downstream of the
terminator cause large changes to the efficiency without causing it to
unbalance.  \cite{chamberlin} Thus balanced termination efficiency is the
norm rather than the exception.

\section{Large Molecules and Glasses}

Large systems are qualitatively different from small ones. \cite{pwa} The
specific heat of all non-crystalline matter in macroscopic quantities -
including biological matter - is proportional to $T$ at low temperatures.
\cite{glass} This behavior is fundamentally incompatible with the linear
vibration of the atoms around sites, and is caused by collective quantum
tunneling of atoms between energetically equivalent ``frustrated''
configurations. \cite{tunnel} It contrasts sharply with the $T^3$ behavior
of crystals with small unit cells. Glasses also exhibit
stretched-exponential time dependence in response to perturbations, i.e.,
of the form $\exp(- A t^\beta)$ with $\beta < 1$, indicating a broad
spectrum of decay rates rather than just one.  They also exhibit memory
effects, such as ``remanence'' in spin glasses \cite{spin} or the
well-known failure of ordinary silica to crystallize without annealing.
This behavior is universal and robust.  All non-crystalline macroscopic
matter exhibits hysteresis, metastability, a broad spectrum of relaxation
times, and memory.

How large a system must be before it can exhibit such behavior is not
known, as the relevant experiments are difficult to perform except on
macroscopic samples, but there are many indications that even medium-sized
proteins have glass-like properties.  Crystals of myoglobin, a protein
with a molecular weight of only 17,000, have linear specific heats at low
temperatures \cite{lineart} and exhibit stretched-exponential response to
photodissociation pulses. \cite{frauen} Denatured proteins refold on a
variety of time scales ranging from nanoseconds to seconds, \cite{doniach}
and amino acids sequences chosen at random will not fold at all.
\cite{wolynes} Permanent misfolding of proteins with molecular weights of
only 30,000 has been implicated in prion diseases. \cite{prusiner} Many
enzymes exhibit hysteresis in their\linebreak\pagebreak

\noindent
catalytic rates. \cite{neet,frieden} The activity of cholesterol oxidase
of {\it Brevibacterium sp.}, a protein with molecular weight 53,000, was
recently shown by fluorescence correlation techniques to have a memory
effect persisting about 1 second under normal conditions at room
temperature. \cite{memory} Other notable examples include wheat germ
hexokinase (mol.\ wt.\ 50,000 \cite{meunier}) with a half-life of 2
minutes, \cite{ricard} rat liver glucokinase (mol.\ wt.\ 52,000
\cite{holroyde}) at 1 minute, \cite{neet} and yeast hexokinase (mol.\ wt.\
50,000) at 1-2 minutes. \cite{shill} Thus RNA polymerase complexes, which
have a molecular weight of 379,000 and are comparable in size to the
largest computer simulations of glasses ever performed, are good
candidates for systems that exhibit glassy behavior.

Glassiness in enzymes is not always easy to observe. The mnemonic effect
in yeast hexokinase occurs when it is preincubated with MgATP and free
Mg$^{2+}$ and the reaction is started with glucose, or preincubated with
glucose and free Mg$^{2+}$ and started with MgATP, but {\it not} if the
enzyme is preincubated with glucose and metal-free ATP and then started
with Mg$^{2+}$. \cite{neet} Mnemonic behavior can be destroyed by
``desensitizing'' the enzyme with contaminants. \cite{meunier} Time scales
can depend on enzyme, substrate, product, activator and effector ligand
concentrations as well as pH, buffers, and temperature.
\cite{neet,shill,peters} Before hysteresis and memory effects were
recognized, early investigators generally adjusted such reaction
conditions until the ``improper'' behavior was eliminated. \cite{neet}

\section{Polymerase States}

While the size of RNA polymerase makes it plausible to expect glassy
behavior on purely theoretical grounds, several direct lines of evidence
indicate that the enzyme exhibits a spectrum of multiconformational,
mnemonic and hysteretic behavior:

\begin{enumerate}

\item Polymerase has a catalytic mode distinct from RNA synthesis, as it
      can cleave the RNA transcript through hydrolysis (rather than
      pyrophosphorolysis, the reverse reaction of RNA synthesis),
      \cite{surratt} with the cleavage reaction requiring Mg$^{2+}$,
      \cite{surratt} being template-dependent, \cite{feng} changing the
      polymerase footprint size, \cite{lee} and stimulated either by GreA
      and GreB proteins \cite{borukhovA,borukhovB} or by high pH
      (8.5-10.0). \cite{orlova} The last effect was discovered
      serendipitously, going unobserved for decades because assay 
      conditions were being optimized to maximize elongation
      rates, which occur at lower pH values (7.8-8.2 
      \cite{chamberlinberg}).\cite{orlova}

\item RNA polymerase mobilities in non-denaturing electrophoresis gels
      show significant and discontinuous variance while bearing nearly
      identical transcripts or identical length transcripts with different
      sequences. \cite{krummelA} These mobility variances are still
      observed if the RNA transcript is first removed by ribonuclease
      digestion. \cite{straney}

\item RNA polymerase ternary complexes vary greatly in their stability and
      mode of binding to DNA (ionic or non-ionic) in a template-dependent
      manner. Some complexes are stable against very high salt
      concentrations ([K$^{+}$] = 1 M), while others (specifically those
      proximal to an upstream palindrome sequence) are salt-sensitive
      (completely dissociating in concentrations as low as 20 mM K$^{+}$).
      However, the salt-sensitive complexes are stabilized by millimolar
      concentrations of Mg$^{2+}$. \cite{arndtB}

\item The size of the RNA polymerase footprint on the DNA template
      measured by ribonuclease digestion is significantly altered even at
      adjacent template positions, suggesting that the enzyme assumes
      different conformations during elongation.  \cite{krummelB}

\item Guanosine tetraphosphate (ppGpp) inhibits the rate of elongation on
      natural DNA templates but not on synthetic dinucleotide polymer
      templates, and does not inhibit elongation by competing with NTP
      binding, but by enhancing pausing. It must therefore bind to
      polymerase and modify its behavior at an unrelated regulatory site
      in an allosteric manner, rather than interfering with the substrate
      binding site. \cite{kingston}

\item The stability of a stalled elongation complex depends on whether the
      polymerase arrives at the stall site via synthesis or 
      pyrophosphorolysis. \cite{rozo}

\item Termination efficiencies are affected by transcribed upstream
      sequences and {\it un}transcribed downstream sequences adjacent to
      the terminator. \cite{tele}

\item Stalling elongating polymerase complexes (via nucleotide starvation)  
      and then restarting them by nucleotide addition perturbs pausing
      patterns 50-60 base pairs downstream. \cite{dissinger}

\item An elongating polymerase's Michaelis constants $K_{S}$ for NTPs vary
      over 500-fold for different DNA template positions, \cite{levin} and
      for different templates, \cite{job} although these effects are not
      observed for synthetic dinucleotide polymer templates. \cite{job}

\item The rate of misincorporation at a single site for which the correct
      NTP is absent is significantly different before and after isolation
      of ternary complexes. \cite{erie}

\item Stalled polymerase gradually ``arrests'' (i.e., is incapable of
      elongating when supplied with NTPs), with the approximate half-time
      for arrest estimated at 5 minutes \cite{arndtB} and 10 minutes 
      \cite{markovtsov} for different DNA templates. The polymerase can
      continue elongating if reactivated by pyrophosphorolysis.
      \cite{arndtB}

\item Even after undergoing arrest, crosslinking experiments show that the
      internal structure of polymerase gradually changes over the course
      of the next hour. \cite{markovtsov}

\item Observations of single elongating RNA polymerase molecules show that
      it has two elongation modes with different intrinsic
      transcription rates and propensities to pause and arrest. 
      \cite{davenport}

\end{enumerate}

The possibility of metastability - through shape memory or the conditional
attachment of factors - is directly relevant to the rate-balance conundrum
because it provides a simple alternative to balanced stochastic branching
that requires no physical miracles.  If, for example, the polymerase
possessed a small number of metastable configurational states and
terminated deterministically depending on which state it was in, then
balanced branching would be a simple, automatic consequence of scrambling
the state populations.

\section{Thermal Activation}

The idea that polymerase memory is potentially relevant to expression
regulation is not new. \cite{job} It is implicit in the work of Goliger et
al \cite{goliger} and Telesnitsky and Chamberlin \cite{tele} and
even explicitly speculated by the latter in print.  However, because of
the experimental evidence supporting the stochastic model of termination
\cite{standard} and the widespread belief - unjustified, in our view -
that proteins equilibrate rapidly, this suggestion generated little
enthusiasm.  A key experiment supporting the stochastic model by Wilson
and von Hippel \cite{wilson} is both historically important and typical,
so it is appropriate that we consider it carefully.

\enlargethispage*{1000pt}

Wilson and von Hippel promoted and stalled RNA polymerase 8 base pairs
upstream of the tR2 terminator hairpin of phage $\lambda$ {\it in vitro},
thermally equilibrated at temperature $T$, and then launched it forward by
adding NTP.  The results are reproduced in Fig.\ 1a. Termination occurred
at sites 7, 8, and 9 base pairs downstream of the beginning of the poly-T
stretch (cf.\ Table~II)  with probabilities $P_7 = N_7/N$, $P_8 = N_8/N$
and $P_9 = N_9/N$.  The data were originally reported as a semilogarithmic
plot of $1/\hat{P} - 1$ against temperature, where $\hat{P}_7 = N_7/N$,
$\hat{P}_8 = N_8/(N-N_7)$ and $\hat{P}_9 = N_9/(N-N_7 - N_8)$.  They
concluded that all three branching probabilities $\hat{P}$ were thermally
activated and had distinctly different activation energies. However, it is
clear from Fig.\ 1a that this conclusion is false.  The three
probabilities $P$ are essentially the same function and are well
characterized by the sum $P = P_7 + P_8 + P_9$, also plotted in Fig.\ 1a. 
This is shown more explicitly in Fig.\ 1b, where the ratios $P_7/P$,
$P_8/P$, and $P_9/P$ are plotted against temperature. The flatness of
these curves shows that the branching ratios among the three sites are
essentially constant and independent of temperature within the error bars
of the experiment.  Note that these fractions are also all of order~1.
Thus the alleged spread in activation energies was an artifact of the
plotting procedure. 

Let us now consider the temperature dependence.  It may be seen
from Fig.\ 1a that $P$ saturates to 1 at 80 $^\circ$C, the temperature at
which Wilson and von Hippel\linebreak\pagebreak

\begin{figure}
\epsfbox{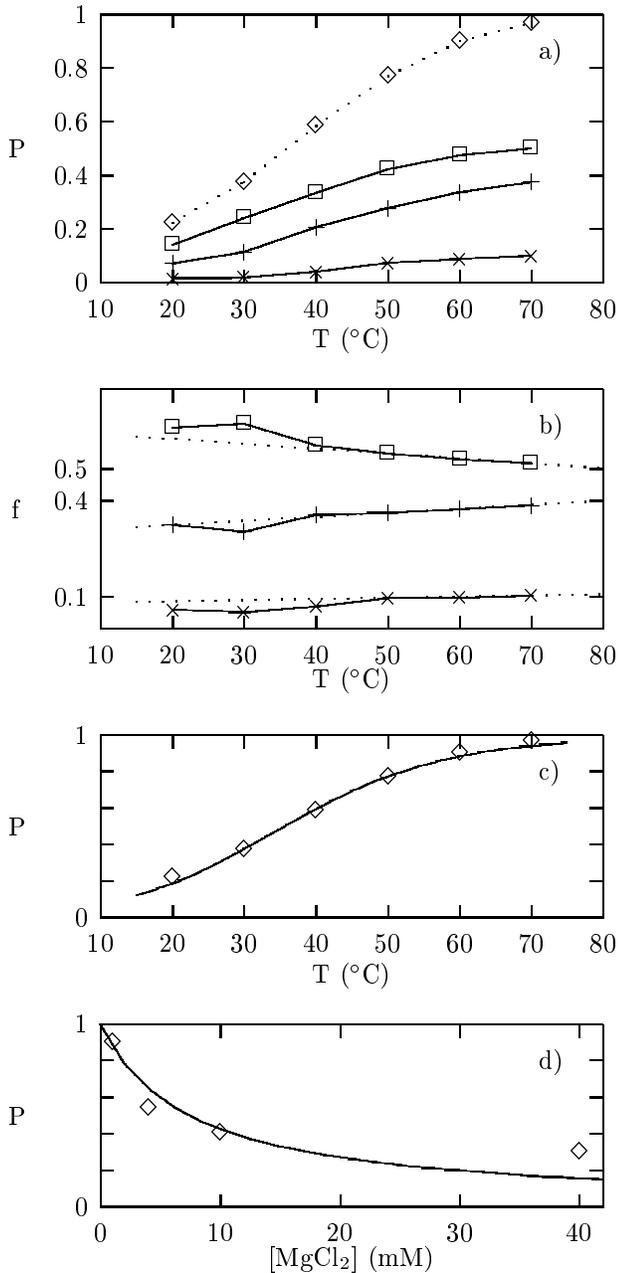}
\caption{a) Temperature dependence of termination probability $P$ for
        phage $\lambda$ terminator tR2 reported by Wilson and von Hippel.
        \cite{wilson} $+$, $\Box$, and $\times$ denote the probabilities
        to terminate 7, 8, and 9 nucleotides downstream from the beginning
        of the poly-T stretch.  The sum is shown as $\Diamond$. b)
        $+$, $\Box$, and $\times$ above divided by $\Diamond$ to make
        a branching fraction $f$. c) Comparison of ionization model
        Eq.~(1) with $\Diamond$ from a).  The ionization energy has been
        fit to $\epsilon_0 = 0.7 eV$ (16 kcals/mole) and the quantity
        $n/M^{3/2}$ adjusted to make the curves match at 30~$^\circ$C. d)
        Prediction of Eqn.~(1) for dependence on  Mg$^{2+}$
	concentration compared with data of Reynolds et al.
	\cite{chamberlin}}
\end{figure} 

\noindent
report that the polymerase ``will not elongate'', i.e., has stopped
working properly.  This suggests that the effect has something to do with
the overall mechanical integrity of the enzyme rather than the termination
process alone.  Guided by this observation we note that the activated
behavior identified by Wilson and von Hippel is actually the formula for
conventional monomolecular chemical equilibrium.  The probability for a
particle of mass $M$ with a binding energy of $E_0$ to be ionized off the
polymerase is

\begin{equation}
P = \frac{1}{1 + Z e^{E_0/k_B T} \; n \; \lambda_{th}^3 }
\; \; \; \; \; \; \; \;
(\lambda_{th} = \sqrt{\frac{2\pi \hbar^2}{M k_B T}} \; )
\; \; ,
\end{equation}

\noindent
where $n$ is the concentration of this component and $Z$ is the change to
the internal partition function that results from binding.  If one makes
the approximation that $\lambda_{th}$ is a slowly-varying function of
temperature and can thus be taken to be constant then this reduces to the
formula with which Wilson and von Hippel fit their data. \cite{wilson} 
That it works may be seen in Fig.~1c, where we plot the total termination
probability from experiment against Eq.~(1) with $E_0 = 0.7 eV$ and $Z$
adjusted to match experiment at T = 30~$^\circ$C.  Thus reinterpreting
this effect as an ionization equilibrium, we may account for the
high-temperature intercept and weak temperature dependence seen in Fig.~1b
in the following way: In addition to the ionization state the polymerase
possesses an internal configurational memory with a number of states of
order~10.  These code for termination at sites 7, 8 or 9.  In the
equilibration step, the polymerase molecules come to thermal equilibrium
and a fraction $P$ of them become ionized.  All of these terminate at one
of the three sites when launched.  The rest read through.

\enlargethispage*{1000pt}

A candidate for the ionizable component is an Mg$^{2+}$ ion.  In their
studies of the effects of ion concentrations on termination efficiency,
Reynolds et al \cite{chamberlin} discovered that Mg$^{2+}$ has the strange
and unique effect of increasing termination efficiency to 100\% for all
terminators studied when reduced below 1 mM.  The Mg$^{2+}$ concentration
in the experiments shown in Fig.~1d was 10 mM. \cite{wilson} Extrapolating
at T = 30~$^\circ$C \cite{schmidt} using Eq.~(1) we obtain, with no
adjustable parameters, the fit to the [MgCl$_2$] dependence found by
Reynolds et al\cite{chamberlin} shown in Fig.~1d.  The quality of this fit
suggests that Mg$^{2+}$ has a special function in regulating
transcription, and that the temperature dependence in Fig.~1a is simply a
thermal binding relation for this ion. This is corroborated by the recent
structural studies of Zhang et al, \cite{crystal} who report that
polymerase crystallized out of 10 mM solution of MgCl$_2$ has a Mg$^{2+}$
ion bound at what appears to be the catalytic site of the enzyme.

There is evidence for more termination channels other than the ionization
of Mg$^{2+}$.  In Fig.~2 we reproduce results of Reynolds et al
\cite{chamberlin} showing that terminator efficiencies tend to saturate at
large Mg$^{2+}$ concentration to values other than zero.  The saturation
values are balanced, and there is an evident tendency of them to cluster.
Both effects are consistent with the polymerase executing an instruction
at the terminator to read through conditionally, even when the ionizable
component is bound, if its memory is appropriately set.  There is
obviously not\linebreak\pagebreak

\begin{figure}
\epsfbox{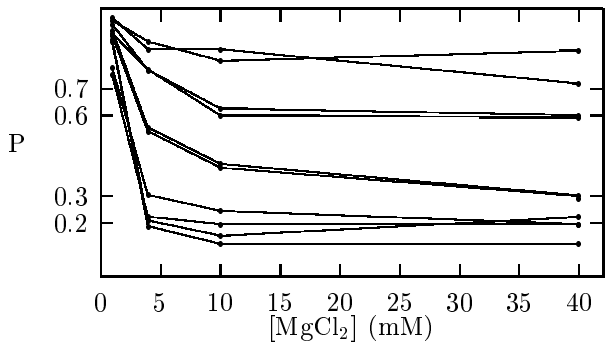}
\caption{Termination efficiency as function of [MgCl$_2$] for 10
terminators, as reported by Reynolds et al. \cite{chamberlin}  The
terminators are, top to bottom at the right edge, {\it RNA I}, {\it T7Te},
{\it rrnB T1}, {\it trp a L126}, {\it trp a}, {\it tR2}, {\it T3Te},{\it
P14}, {\it tonB t}, and {\it trp a L153}.}
\end{figure}

\noindent
enough data here to draw such a conclusion, however. We note that Reynolds
et al \cite{chamberlin} also found order-1 effects on the termination
efficiency from Cl$^-$ and K$^+$, although with the opposite sign.  The
function of these ions is not yet known.

\section{Antitermination}

What experiments can detect internal memory?  In general, one would look
for cases in which polymerase acts differently under apparently identical
conditions, suggesting an internal control mechanism of some kind.  Such
thinking motivates the following hypothetical experiment: one constructs a
template with promoter $P$ followed by two identical terminators and
flanking DNA sequences in succession.  If termination is stochastic, then
the branching ratio at T$_2$ will be the same as that at T$_1$.  If
termination is deterministic and hysteretic, then the branching ratios
will be different, depending on details.  A passive termination at T$_1$
would result in no termination at T$_2$, since the polymerase that reads
through has been ``polarized'', i.e., selected for the memory setting that
codes for read-through.  An active termination at T$_1$ would reprogram
the memory there and cause a termination probability at T$_2$ different
from that of T$_1$ but not necessarily zero.  Variations of this design,
e.g., adding more terminators, combining different terminators, changing
their order, etc., could, in principle, answer more sophisticated
questions, such as whether and how polymerase is reprogrammed in active
read-through and whether non-equilibrium effects are important. 

A few such experiments have already been performed on DNA
templates containing antiterminators (sequences upstream of
terminators that reduce termination efficiencies) and are thus less
general than one would like, but they strongly support the idea
of polymerase memory.  There is indirect evidence in the case of
N-antitermina-\linebreak tion of phage $\lambda$, the case most studied,
that the
memory is a physical attachment of the transcribed mRNA to the\linebreak


\begin{table}
\begin{tabular}{|l|c|c|}
\multicolumn{1}{|c|}{Sequence} & T7Te & trp a \\
\hline
{\tt
\underline{AATTGT}GAGCGGATA\underline{ACAATT}TCACACAGGAAACAGGGAA}
& 61 & 99 \\
{\tt
\underline{AATTGT}GAGCGGATA\underline{ACAATT}TCACACAGGAAACAGAA..}
& 51 & 52 \\
{\tt
\underline{AATTGT}GAGCGGATA\underline{ACAATT}TCACACAGGAA...}
& 73 & 99 \\
{\tt
\underline{AATTGT}GAGCGGATA\underline{ACAATT}TCACGGAA...}
& 45 & 99 \\
{\tt
\underline{AATTGT}GAGCGGATA\underline{ACAATT}TCAGGAA...}
& 71 & 99 \\
{\tt
\underline{AATTGT}GAGCGGATA\underline{ACAATT}TCGGAA...}
& 75 & 66 \\
{\tt
\underline{AATTGT}GAGCGGATAGGAA...}
& 88 & 75 \\
\multicolumn{1}{|c|}{No Antiterminator} & 99 & 80 \\
\end{tabular}
\caption{Sequences and corresponding termination probabilities at
downstream {\it T7Te} and {\it trp a} for modified {\it lac}
antiterminators reported by Telesnitsky and Chamberlin. \cite{tele}}
\end{table}

\begin{table}
\begin{tabular}{|l|c|c|}
\multicolumn{1}{|c|}{Sequence} & oop t & rpoC t \\
\hline 
{\tt
\underline{AAAT}CTGATA\underline{ATTT}TGCCAATGTTGTACG\underline{GAATTC}}
& 37 & 22 \\
{\tt
\underline{AAAT}CTGATA\underline{ATTT}TGCCAATGTTGG\underline{GAATTC}...}
& 45 & 17 \\
{\tt
\underline{AAAT}CTGATA\underline{ATTT}TGCCAATGTTG\underline{GAATTC}...}
& 31 & 19 \\
{\tt
\underline{AAAT}CTGATA\underline{ATTT}TGCCAATG\underline{GAATTC}...}
& 29 & 16 \\
{\tt
\underline{AAAT}CTGATA\underline{ATTT}TGCCG\underline{GAATTC}...}
& 25 & 18 \\
{\tt
\underline{AAAT}CTGATA\underline{ATTT}G\underline{GAATTC}...}
& 17 & 20 \\
{\tt
A\underline{AAT}CTGATA\underline{ATT}G\underline{GAATTC}...}
& 15 & 22 \\
{\tt
AA\underline{AT}CTGATA\underline{AT}G\underline{GAATTC}...}
& 11 & 20 \\
{\tt
AAA\underline{T}CTGATA\underline{A}G\underline{GAATTC}...}
& 19 & 21 \\
{\tt
AAATCG\underline{GAATTC}...}
& 20 & 16 \\
\end{tabular}
\caption{Antiterminator sequences constructed by Goliger et al
\cite{goliger} from a promoter from phage 82, together with the
readthrough probabilities {\it in vitro} for downstream terminators
{\it oop t} and {\it rpoC t}.  Note that these terminators are not in
series. The underlined sequence on the right is the EcoRI linker.}
\end{table}

\noindent
polymerase to form a loop.  \cite{nodwell} There is also evidence
that it is not true generally.  \cite{tele}

In 1989 Telesnitsky and Chamberlin \cite{tele} reported memory effects
associated with the {\it lac} antiterminator found just downstream of the
{\it Ptac} promoter in {\it E. coli}.  Their key result is reproduced in
Table V. Insertion of {\it lac} 353 nucleotides upstream of the terminator
makes different order-1 modifications to the termination efficiences of
{\it T7Te} phage and {\it trp a}.  The antiterminator contains a
palindrome, and the antitermination effect is sensitive to modifications
of the downstream 15-base-pair sequence.  3 copies of {\it T7Te} placed in
tandem downstream of {\it lac} showed that the antitermination effect is
partially remembered through multiple terminators: the efficiencies were
44\%, 60\%, and 90\%, but without the antiterminator they were 90\%,
$>$90\%, and $>$90\%.

In another experiment {\it in vitro} reported in 1989, Goliger et al
\cite{goliger} found that the {\it E. coli} terminator {\it rpoC t} and
phage terminators {\it oop t} and {\it t$_{82}$} were strongly
antiterminated by a sequence they constructed accidentally.  Their key
result is reproduced in Table VI.  A phage 82 promoter was fused onto a
sequence containing either {\it rpoC t} alone or {\it oop
t}\linebreak\pagebreak

\enlargethispage*{1000pt}

\begin{table}
\begin{tabular}{|c|c|}
Sequence & Name \\
\hline
{\tt
GAGCGCGGCGG\underline{GTTCA}GGA\underline{TGAAC}GGCAATGCTGCTCATTAGC}
& putL \\
{\tt
GCGTG\underline{GTCA}AGGA\underline{TGAC}TGTCAATGGTGCACGATAAAAACCCA}
& putR \\
\end{tabular}
\caption{Antitermination sequences  {\it putL} and {\it putR} from the
Hong Kong phage HK022.\cite{hkphage}}
\end{table}

\noindent
and {\it rpoC t} in tandem using the EcoRI linker sequence {\tt GGAATTC}. 
This resulted in unexpected antitermination {\it in vitro} of both
terminators, but of different sizes that depended sensitively on the
insertion point.  The read-through effects in the tandem experiments were
unfortunately poorly documented.  One can see from Table V that the phage
terminator responded more strongly in this experiment than did {\it rpoC
t}. However, the reverse was the case in another experiment in which the
antiterminator was a portion of the 6S RNA gene downstream of a phage
$\lambda$ pR$'$ promoter, and in which factor {\it NusA} was present.  As
a control, this latter experiment was rerun with the phage terminator {\it
t$_{82}$}, which terminated at greater than 98\% in all cases, seemingly
immune to antitermination. 

King et al \cite{king} reported in 1996 that the {\it putL} and {\it putR}
antitermination sequences of the Hong Kong phage HK022, \cite{hkphage}
shown in Table~VII, caused downstream readthrough of a triple terminator
consisting of tR$'$ from phage $\lambda$ followed by the strong {\it E.
coli} ribosome operon terminators {\it rrn B t$_1$} and {\it rrn B t$_2$}. 
This effect was sensitive to the choice of promoter.  When {\it putL} was
inserted between the {\it Ptac} promoter and the triple terminator 284
nucleotides downstream and studied {\it in vivo} the termination
probability was 50\%.  Substituting the phage $\lambda$ P$_L$ promoter for
{\it Ptac} under the same conditions resulted in complete readthrough
(though with wide error bars). When this experiment was repeated {\it in
vitro} the antitermination effect was found to be smaller and to persist
through all three terminators. The read-through probabilities at tR$'$
were 34\% and 31\% for promotion by P$_L$ and {\it Ptac}, respectively,
but 57\% and 27\% for {\it rrnB t$_1$} and 76\% and 40\% for {\it rrnB
t$_2$}.  This result is incompatible with statistical termination, for
both the antitermination effect itself and the changes resulting from
switching promoters are order-1 effects that do not add.  They also
reported that reduced Mg$^{2+}$ concentration destroys the antitermination
effect.

\section{Conclusion}

In summary we find that the theory of stochastic termination, which
requires natural selection to engineer a physical miracle of balanced
rates, is flawed, but that there is ample evidence of a sophisticated and
as-yet poorly understood regulatory system in RNA polymerase involving
hysteresis, metastability, and long-term configurational memory, all
robust phenomena in inanimate matter.  On this basis we
predict that branching ratios of\linebreak\pagebreak 

\noindent
identical terminators in series will differ by order-1 amounts very
generally - specifically in the absence of looping.  We propose that the
confusion surrounding the existence of polymerase memory is symptomatic of
the larger problem that measurement of physical activity on the length and
time scales appropriate to life has thus far been impossible, and that
overcoming this problem should be one of the high-priority goals of modern
nanoscience. 

This work was supported primarily by NASA Collaborative Agreement NCC
2-794.  SL would like to acknowledge informative discussions with Brian
Ring.  RBL wishes to express special thanks to the organizers of the
Keystone conferences, particularly R. Craig, for encouraging
interdisciplinary science.  We also wish to thank the Institute for
Complex Adaptive Matter at Los Alamos, the Brown-Botstein group at
Stanford for numerous stimulating discussions, and H. Frauenfelder, J. W.
Roberts, D. Botstein and M. J. Chamberlin for a critical reading of the
manuscript.

\end{document}